\documentclass [12pt]{article}
\usepackage {graphicx}
\usepackage[american]{babel}

\begin{document}

\title{On the road towards multidimensional tori.}

\author{Kuznetsov A.P., Sataev I.R., Turukina L.V.}

\maketitle
\begin{center}
\textit{Kotel'nikov's Institute of Radio-Engineering and Electronics of RAS, Saratov Branch,\\
Zelenaya 38, Saratov, 410019, Russian Federation.\\}
\end{center}

\begin{abstract}
The problem of persistence of four-frequency tori is considered in
models represented by the coupled periodically driven
self-oscillators. We show that the adding the third oscillator
gives rise to destruction of the three-frequency tori, with
appearance of regions of either chaotic attractors or
four-frequency tori. As the coupling strength decreases, the
four-frequency tori dominate, and the amplitude threshold of their
occurrence vanishes. Also, for three oscillators, a domain of
complete synchronization of the system by the external driving can
disappear.
\end{abstract}


\section{INTRODUCTION}

A problem of effect of external driving in systems of mutually
coupled self-oscillators is a fundamental part of the theory of
synchronization important for a variety of applications \cite{b1},
\cite{b2}. It is a part of a field related to birth and
synchronization of multi-frequency quasi-periodic oscillations. In
particular, it goes back to the concept of the onset of turbulence
advanced by Landau and Hopf, which is based on successive
appearance of growing number of quasi-periodic components of the
motion with increase of the Reynolds number \cite{b3}. As known,
further developments of nonlinear dynamics and bifurcation theory
have made this hypothesis doubtful in some respects. According to
Ruelle and Takens \cite{b4}, in systems with multiple oscillatory
components arising successively due to a sequence of bifurcations,
the quasi-periodic motions associated with torus attractors cannot
occur as typical at a number of frequency components larger than
three; instead a strange attractor should appear. On the other
hand, computations for model systems undertaken by several authors
do not quite agree with this statement: in certain situations tori
of sufficiently high dimension seem persisting and observable \cite{b5}.
Thus, the question of how the transition to chaos in
multidimensional systems develops, and what is the role of the
number of oscillatory components, or degrees of freedom, involved
in the dynamics, deserve further studies. The phase approximation
whereby dynamics of coupled periodically driven
self-oscillators is investigated appears to be
effective and illuminating in the study of
this problem \cite{b5}. In this paper, in the framework of the
phase equations we discuss transformation of the regions of the
different regimes in the parameter space when the number of
driven oscillators increases. Thus, we will be interested in the
question of how the transition to high-dimensional tori and chaos
depends on the number of oscillators in the system. Also, we show
that an increase in the number of oscillators up to three leads to
a new effect - a domain of complete synchronization may disappear.

\section{PHASE EQUATIONS}

Let us consider a system of three dissipative coupled van der Pol
oscillators driven by an external periodic force (Fig.1):
\begin{equation}
\label{eq1}
\setlength\arraycolsep{2pt}
\begin{array}{l}
  \ddot{x} - (\lambda - x^{2}) \dot{x} + x + \mu (\dot{x} - \dot{y}) = B \sin \omega t, \\
  \\
  \ddot{y} - (\lambda - y^{2}) \dot{y} + (1 + \Delta_{1}) y + \mu (\dot{y} - \dot{x})
  + \mu (\dot{y} - \dot{z}) = 0, \\
  \\
  \ddot{z} - (\lambda - z^{2}) \dot{z} + (1 + \Delta_{2}) z + \mu
  (\dot{z} - \dot{y}) = 0.
\end{array}
\end{equation}
Here $\lambda$ is a parameter responsible for the Andronov -- Hopf
bifurcation in the independent oscillators, $\Delta_{1,2}$ are
frequency detuning of the second and third oscillators from the first
oscillator, $\mu$ is the coupling coefficient, $B$ is an amplitude
of the external periodic force, $\omega$ is its frequency. The
frequency of the first oscillator is assumed to be
normalized to unity.

\begin{figure}[!ht]
\centerline{
\includegraphics[scale=0.75]{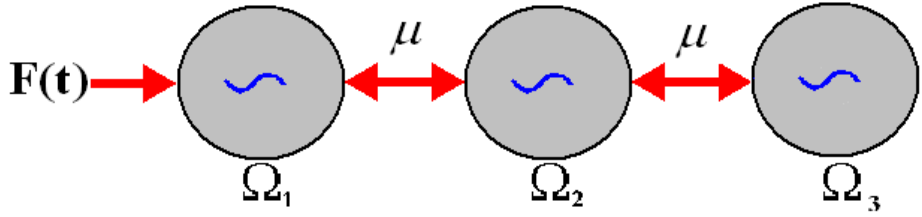}\\}
\caption{Schematic representation of a system of three dissipative
coupled oscillators driven by an external periodic force.}
\end{figure}

The system (\ref{eq1}) may be analyzed in terms of complex
amplitudes (quasiharmonic approximation) if the parameter
$\lambda$ is small. For this, we set
\begin{equation}
\label{eq2} \setlength\arraycolsep{2pt}
\begin{array}{c}
  x = a(t) e^{i \omega t} + a^{\ast} (t) e^{- i \omega t}, \quad
  y = b(t) e^{i \omega t} + b^{\ast} (t) e^{- i \omega t}, \\
   \\
  z = c(t) e^{i \omega t} + c^{\ast} (t) e^{- i \omega t}.
\end{array}
\end{equation}
Here, $a$, $b$, $c$ are the complex amplitudes of the oscillators
slow varying in time scale of the basic oscillations with unit
frequency. We use the additional conditions traditional for this
method
\begin{equation}
\label{eq3} \setlength\arraycolsep{2pt}
\begin{array}{c}
 \dot{a}(t) e^{i \omega t} + \dot{a}^{\ast} (t) e^{- i \omega t}=0, \quad
 \dot{b}(t) e^{i \omega t} + \dot{b}^{\ast} (t) e^{- i \omega t}=0, \\
 \\
 \dot{c}(t) e^{i \omega t} + \dot{c}^{\ast} (t) e^{- i \omega t}=0.
\end{array}
\end{equation}

Then, we substitute (\ref{eq2}) and (\ref{eq3}) in (\ref{eq1}),
multiply the equations by factor $e^{- i \omega t}$, and perform
averaging over a period of the basic oscillations. Assuming
$\omega \approx 1$ the resulting equations for complex
amplitudes are as follows
\begin{equation}
\label{eq4}
\setlength\arraycolsep{2pt}
\begin{array}{rl}
  2 \dot{a}= & \lambda a - |a|^{2} a - 2 i \Omega a - \mu (a - b) - \displaystyle\frac{b}{4},\\
  \\
  2 \dot{b}= & \lambda b - |b|^{2} b - i (2 \Omega - \Delta_{1}) b - \mu (b - a) - \mu (b - c), \\
  \\
  2 \dot{c}= & \lambda c - |c|^{2} c - i (2 \Omega - \Delta_{2}) c - \mu (c -
  b),
\end{array}
\end{equation}
where $\Omega = \omega - 1$ is detuning parameter of the external
driving force with respect to the first oscillator. Note that in the
equations (\ref{eq4}) the parameter $\lambda$ may be eliminated via
variable change, and then the other parameters will be
renormalized with the factor $\lambda^{-1}$ \cite{b5}. Let us set
$a = R e^{i \psi_{1}}$, $b = r e^{i \psi_{2}}$, $c = v e^{i
\psi_{3}}$, where $R$, $r$, $v$ are real amplitudes, and
$\psi_{1,2,3}$ are phases of the oscillators. Following the Refs.
\cite{b1}, \cite{b4}, \cite{b5}, we assume that the motion takes
place close to the unperturbed limit cycles of the oscillators,
and set $R \approx 1$, $r \approx 1$, $v \approx 1$. Then, the
phases have to satisfy the equations
\begin{equation}
\label{eq5}
\setlength\arraycolsep{2pt}
\begin{array}{rl}
  \dot{\psi}_{1}= & - \Omega - \displaystyle\frac{1}{2} \mu \sin (\psi_{1} - \psi_{2})
  + b \sin \psi_{1},\\
  & \\
  \dot{\psi}_{2}= & - \Omega + \displaystyle\frac{1}{2} \Delta_{1} +
  \displaystyle\frac{1}{2} \mu \sin (\psi_{1} - \psi_{2})
  -  \displaystyle\frac{1}{2} \mu \sin (\psi_{2} - \psi_{3}), \\
   & \\
  \dot{\psi}_{3}= & - \Omega + \displaystyle\frac{1}{2} \Delta_{2} +
  \displaystyle\frac{1}{2} \mu \sin (\psi_{2} -
  \psi_{3}).
\end{array}
\end{equation}

Further we shall use the relative phases of the oscillators
$\theta = \psi_{1} - \psi_{2}$ and $\varphi = \psi_{2} -
\psi_{3}$. Equations (\ref{eq5}) generalize the equations obtained
in Ref.\cite{b4} to the case of three driven oscillators.

\section{SYNCHRONIZATION OF TWO- \\
FREQUENCY OSCILLATIONS}

First, let us turn to the case of two oscillators. Then phase
equation are
\begin{equation}
\label{eq6} \setlength\arraycolsep{2pt}
\begin{array}{rl}
  \dot{\psi}_{1}= & - \Omega - \displaystyle\frac{1}{2} \mu \sin \theta
  + b \sin \psi_{1},\\
  & \\
  \dot{\psi}_{2}= & - \Omega + \displaystyle\frac{1}{2} \Delta_{1} +
  \displaystyle\frac{1}{2} \mu \sin \theta.
\end{array}
\end{equation}

We fix the "internal" parameters of the system $\Delta_{1} = 0.3$
and $\mu = 0.4$. Obviously, in absence of the external driving $b
= 0$, the condition of mutual synchronization of the subsystems is
$\mu > \displaystyle\frac{\Delta_{1}}{2}$ \cite{b1}. Hence, the
above selected parameter values correspond to the regime of the mode
locking for the unforced system. The external signal applied can
destroy this regime.

To visualize the domains of different dynamical regimes the
results of parameter scan are represented in Fig.2a for the
($\Omega$, $b$) parameter plane. Shown are the region of complete
synchronization of the phases of oscillators by the external
driving $P$ and regions of the quasiperiodical regimes, namely,
two-frequency tori $T_{2}$ and three-frequency tori $T_{3}$. For
each parameter value we perform computations to reach a sustained
dynamical regime of the model (\ref{eq6}), and then evaluate two
Lyapunov exponents, $\Lambda_{1}$ and $\Lambda_{2}$, using a
standard Bennettin algorithm. The nature of the regimes is
determined as follows. The case $\Lambda_{1} < 0$, $\Lambda_{2} <
0 $ corresponds to a complete phase locking of the oscillator
phases by the external force. The cases $\Lambda_{1} = 0$,
$\Lambda_{2} < 0$ and $\Lambda_{1} = \Lambda_{2} =
0$\footnote{Checking the zero Lyapunov exponents was carried out
with an accuracy of $10^{-3}$. The accuracy of calculating the
main Lyapunov exponents was at least $10^{-4}$.} are interpreted as
associated with quasiperiodic regimes of two-frequency tori and
three-frequency tori, respectively.

It should be noted, that numerical tests are inevitably crude. In
particular, some of the points labelled as three-frequency
quasiperiodic may well be high-order resonance two-frequency
quasiperiodic, and four-frequency quasiperiodic may well be
three-frequency quasiperiodic with high-order relation.
Furthermore, such a Lyapunov classification labels as
quasiperiodic some points, at which the Lyapunov exponent may
vanish, for example, bifurcation points, but this situations
usually take place only at the borders of domains of stable
dynamical regimes and do not influence the overall picture.

The two-frequency regimes $T_{2}$ correspond to a partial locking
of the oscillatory system by the external periodic force, while
the relative phase oscillates near some stationary value. Observe
that these regimes dominate at small driving amplitudes, that is,
they occupy a largest area on the parameter plane there. At large
amplitudes, again the two-frequency tori occur, but now the first
oscillator is partially locked. In the intermediate domain of
moderate amplitudes, the three-frequency tori $T_{3}$ are
observed, and the region of their occurrence is pierced by narrow
tongues of higher order resonance two-frequency tori.

\section{SYNCHRONIZATION OF THREE- \\
FREQUENCY OSCILLATIONS}

Now, let us add the third oscillator into the model system, and turn
to the complete set of equations (\ref{eq5}). We keep the same
parameter values $\Delta_{1} = 0.3$ and $\mu = 0.4$, and set
additionally $\Delta_{2} = 1$. In the absence of the external
force, presence of the third oscillator gives rise to regime of
beats at these parameter values. In other words, the third
oscillator destroys the mutual mode-locking of the first and
second oscillators. As a result, essential complication of the
arrangement of domains in the parameter plane ($\Omega$, $b$) is
observed.

Fig.2b shows the parameter plane chart for the system of three
coupled driven oscillators. Now, the picture is drawn using
numerical analysis of three Lyapunov exponents. New formations are
observed on the diagram. These are domains of four-frequency tori
$T_{4}$, where $\Lambda_{1} = \Lambda_{2} = \Lambda_{3} = 0$ (up
to numerical errors), and domains of chaos $C$, with one positive
exponent, that is $\Lambda_{1} > 0$, $\Lambda_{2} < 0$,
$\Lambda_{3} < 0$.

\begin{figure}[!ht]
\centerline{
\includegraphics[scale=0.70]{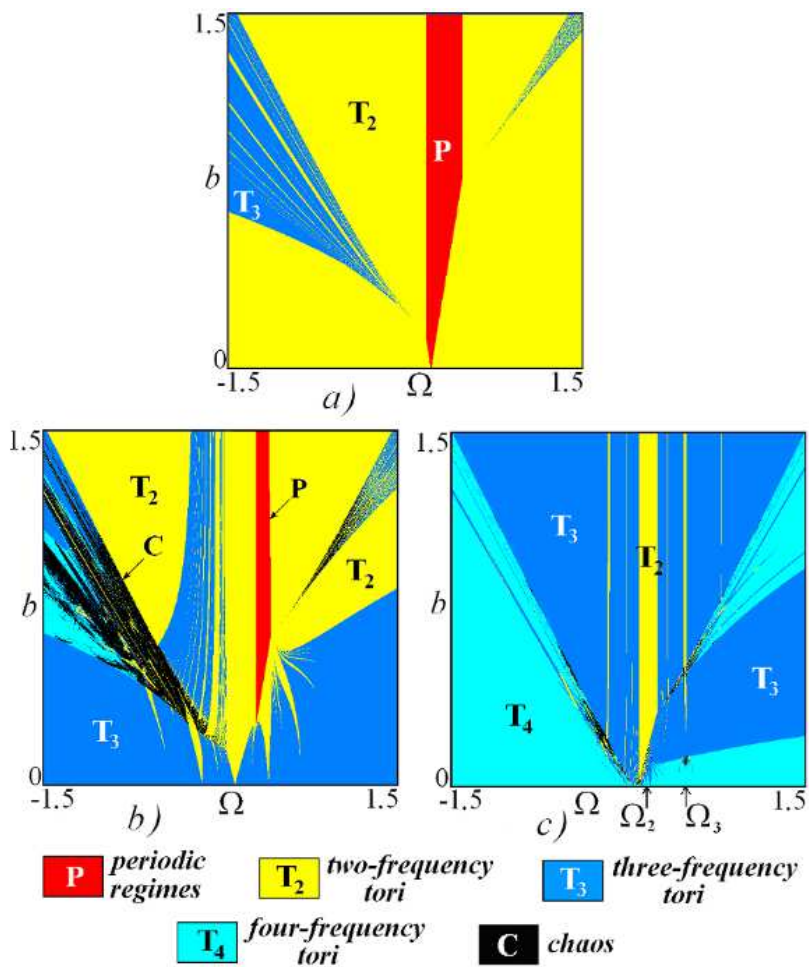}\\}
\caption{Charts of Lyapunov exponents for the model systems of two
(a) and three (b and c) driven oscillators. Parameter values are
$\Delta_{1} = 0.3$, $\Delta_{2}=1$, $\mu = 0.4$ (a), $\mu = 0.4$
(b), $\mu = 015$ (c). The color coding is indicated in the bottom
part of the figure. The marks $\Omega_{2}$ and $\Omega_{3}$
indicate frequencies of the second and the third oscillators respectively.}
\end{figure}

One can see that the domain of complete locking of all oscillators
by the external driving force $P$ decreases notably in width.
Besides, a finite amplitude threshold of the complete
synchronization appears.

The regions of three-frequency tori disappear, and instead of them
the four-frequency tori and chaos arise. Besides, a finite
amplitude threshold of the complete synchronization appears. At small amplitude
three-frequency tori replace the two-frequency quasiperiodicity.

It is interesting to discuss the picture in the context of the Ruelle --
Takens hypothesis. Adding the third oscillator obviously facilitates
destruction of the three-frequency tori, but it can result in appearance
either of chaos, or of the four-frequency tori. The domains of occurrence of
these types of attractors coexist on the parameter plane. Both
quasiperiodicity and chaos look typical, no one prevails. So, we do not
observe an extrusion of the tori by the chaotic regimes, as one could
expect, perhaps, basing on the Ruelle -- Takens assertions.

The winding-number chart of two-frequency tori shown in Fig.3
demonstrates the complexity of the observed picture.

\begin{figure}[!ht]
\centerline{
\includegraphics[scale=0.65]{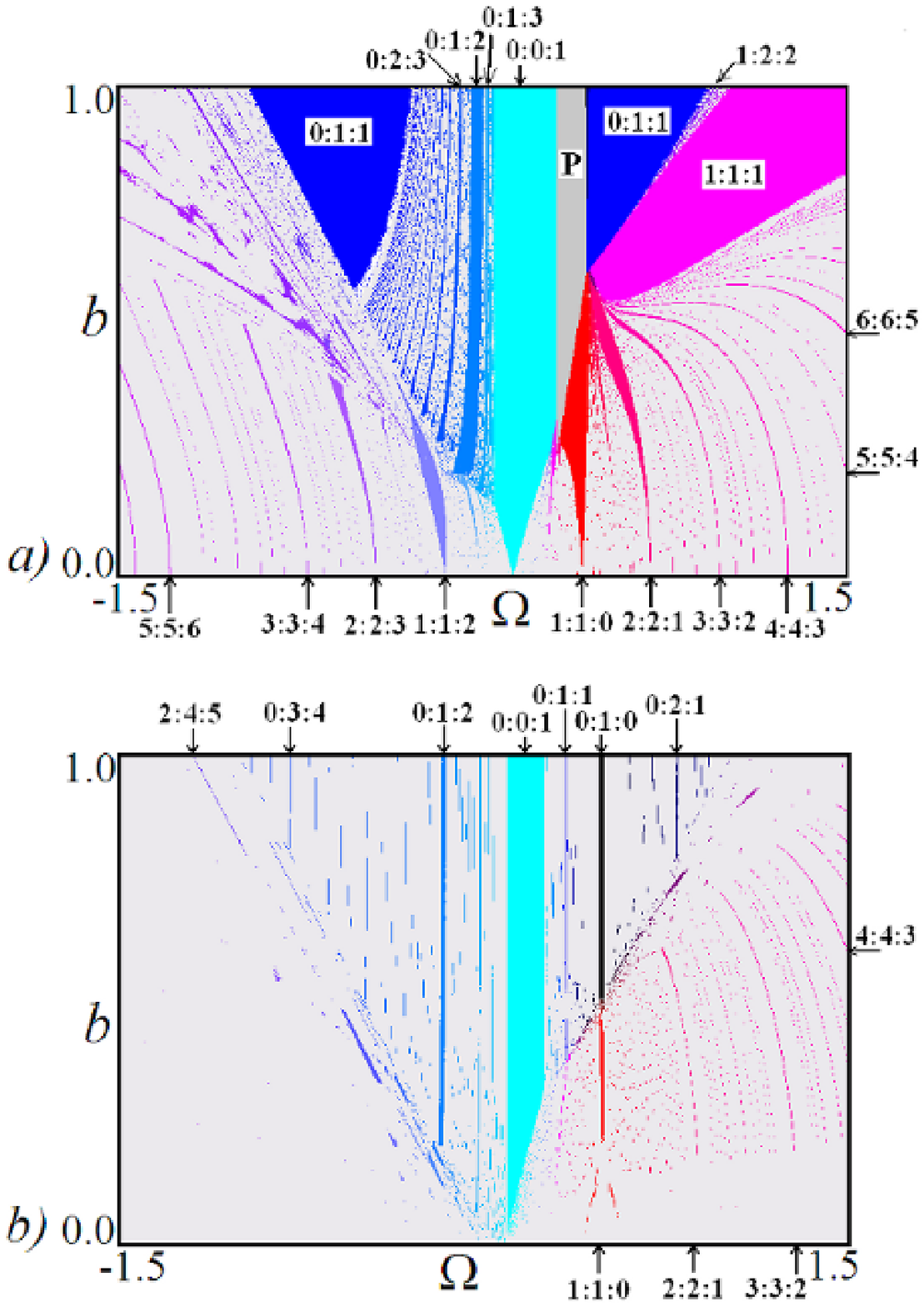}\\}
\caption{A chart of two-frequency tori for the system (\ref{eq5}).
Parameters are corresponded to Figs.1b and c.}
\end{figure}

In terms of phase equations (\ref{eq5}), a two-frequency torus
corresponds to a closed invariant curve in the phase box
($\psi_{1}$, $\psi_{2}$, $\psi_{3}$), with boundaries (0, 2$\pi$).
The parameter scan in Fig.3 is obtained as follows. For each
parameter value we determine via numerical simulations a set of
numbers $p$, $q$, $r$, which are essentially the numbers of
intersections of the invariant curve with the edge surfaces of the
phase box (with exclusion of returns). These numbers determine
what we call the winding number of the invariant curve: $w = p : q
: r$. Then, each point is attributed with a definite color
depending on the winding number we got. The light gray tone
designates regimes, which do not have a definite quantifier in the
frame of the procedure, like the three-frequency tori,
four-frequency tori, or chaos. The domains of main resonances are
marked with subscriptions in the respective areas of the chart in
Fig.3. Note that the regimes of type $w = 0 : q : r$ correspond to
partial locking of the first oscillator by the external force, and
the regime $w = 1:1:1$ corresponds to synchronization of all
oscillators as the unified system.

Now let us consider transformation of the picture under decrease
of the coupling between the oscillators up to the value $\mu =
0.15$. The respective diagram is shown in Fig.2c. In this case,
the four-frequency tori dominate, and they effectively extrude the
chaotic regimes, which occupy now only some extremely narrow
strip-like areas placed mainly along the borders of
three-frequency tori domains. The four-frequency tori dominate as
well at small amplitudes, so, the amplitude threshold for them
vanishes. Additionally, there is a small chaotic region in a
neighborhood of the frequency of the third oscillator, which has,
however, a non-zero amplitude threshold.

A characteristic feature of the Fig.2c is disappearance of the
regime of the complete synchronization $P$. Instead, a narrow
strip of the two-frequency tori remains. The winding-number chart
for this case shown in Fig.3b illustrates this effect. One can see
the regimes with winding number $w = 0 : 0 : 1$ instead of the
region of complete synchronization. This regime corresponds to
partial locking of the first and second oscillator by the external
force. This phase of the third oscillator demonstrates unlimited
growth. Thus the system of three oscillators can be configured so
that complete synchronization is impossible. This is important
difference from the case of two oscillators.

In this paper we have considered the chain of three oscillators
with external force effecting the outermost one. Otherwise, there
are many possibilities for coupling and forcing: we may take the
ring of oscillators or affect the oscillators simultaneously, etc.
It may be conjectured that 4D-tori would persist for either configuration,
but what concerns the possibility of absence of complete synchronization,
the answer may depend on concrete configuration of coupling and
external forcing.

\section{CONCLUSION}

A passage from two-mode to three-mode self-oscillatory system
driven by periodic external signal gives rise to destruction of
three-frequency tori with formation of regions of four-frequency
tori and chaos. At moderate coupling parameter values neither of these regimes
dominates. Decrease of coupling leads to almost complete extrusion
of chaos by the four-frequency quasiperiodic regimes. Besides,
such regimes dominate at small amplitudes of external driving. New
effects for the case of three oscillators is that at sufficiently
small coupling the regime of the complete synchronization of three
oscillators by the external driving disappear.

The work has been supported by RFBR, grant No 09-02-00426.

\end{document}